\documentstyle[subeqn,preprint,aps,epsfig]{revtex}
\begin{document}
 
\def\real{I\negthinspace R}
\def\zed{Z\hskip -3mm Z }
\def\half{\textstyle{1\over2}}
\def\quarter{\textstyle{1\over4}}
\newcommand{\be}{\begin{equation}}
\newcommand{\ee}{\end{equation}}
\newcommand{\bea}{\begin{eqnarray}}
\newcommand{\eea}{\end{eqnarray}}
\preprint{DTP/98/1, gr-qc/9802015}
\draft
\tighten
\renewcommand{\topfraction}{0.8}
%%%%%%%%%%%%%%%%% HERE comment out next 2 lines
%@twocolumnfalse\endcsname
 
\title{DILATONIC GLOBAL STRINGS }
\author{Owen Dando and Ruth Gregory}
\address{Centre for Particle Theory, Department of Mathematical Sciences\\
South Road, Durham, DH1 3LE}
\date{\today}
\maketitle
\begin{abstract}
We examine the field equations of a self-gravitating global string
in low energy superstring gravity, allowing for an arbitrary coupling
of the global string to the dilaton. Massive and massless dilatons are
considered. For the massive dilaton the spacetime is similar to the
recently discovered non-singular time-dependent Einstein self-gravitating
global string, but the massless dilaton generically gives a singular
spacetime, even allowing for time-dependence.
We also demonstrate a time-dependent non-singular string/anti-string
configuration, in which the string pair causes a compactification
of two of the spatial dimensions, albeit on a very large scale.
\end{abstract}
 
%%%%%%%%%%%%%%%%% HERE comment out next  line
%\vskip2pc]
 
%%%%%%%%%%%%%%%%%%%%%%%%%%%%%%%%%%%%%%%%%%%%%%%%%%%%%%%%%%%%%%%%%%%%%% 

\section{Introduction}

Topological defects crop up widely in physics; in cosmology they have been 
put forward as a possible source for
the density perturbations which seeded galaxy formation \cite{BSV}. Phase 
transitions in the early universe can give rise to various forms of topological 
defect. A defect is a discontinuity in the vacuum and is classified 
according to the topology of the vacuum manifold of the field theory model being
used. Disconnected vacuum manifolds give domain walls, non-simply connected 
manifolds, strings, and vacuum manifolds with non-contractible 
spheres give monopoles. Strings and monopoles 
can be further divided into local and global defects depending on the nature of 
the symmetry broken. 
Local defects are formed when the symmetry broken is gauged, and 
the Higgs mechanism typically removes any Goldstone bosons, 
meaning that the defect has a well-defined core and finite 
energy per unit defect area. Global defects on the other hand 
(with the exception of the domain wall which does not result 
from the breaking of a continuous symmetry) typically have a 
residual Goldstone field in the Lagrangian, which translates 
to a power law fall-off in the core energy density and a 
divergent total energy. More unusual defects, such as semi-local
strings \cite{VA} also exist, but we do not consider these here.

Given this difference between local 
and global defects, we might expect them to have
significantly different behaviour, and this is particularly evident in 
their coupling to gravity. Whilst local strings \cite{LSTR} and monopoles 
\cite{LMON} are well-behaved and asymptote flat or locally flat spacetimes, 
global strings \cite{GSTR,G} and monopoles \cite{BV,HL} have strong effects at 
large distances. Indeed, the spacetime of a global string was for
some time thought to be 
singular \cite{GSTR}. In fact, the spacetime is time-dependent 
\cite{G} with a de-Sitter expansion along the spatial extent of the defect.
The spacetime is also characterised by an event horizon at a finite
distance from the string core, although this horizon appears to be
unstable to perturbations \cite{INSTAB}.
Of course, this work has all been performed within the 
context of general relativity. 
At sufficiently high energy scales it seems likely that gravity is not 
given by the Einstein action, but becomes modified, for example, by superstring
terms which are scalar-tensor in nature \cite{LESG}. 
Low energy string gravity is reminiscent of the scalar-tensor
theories of Jordan, Brans and Dicke 
\cite{JBD}, and is equivalent to 
Brans-Dicke theory for a particular value of the Brans-Dicke parameter : 
$\omega=-1$. The implications of Brans-Dicke gravity for defects have
been explored for local strings \cite{GO} and global monopoles \cite{BR}.
However, since the Brans-Dicke parameter is known to be constrained by
$\omega > 500$ \cite{REAS}, we are more interested in exploring the 
low energy superstring action, particularly with a massive dilaton, although
massless dilatons will provide a comparison with Brans-Dicke results.
Moreover, some recent superstring inspired models of inflation \cite{AR}
appear to
allow for cosmic string defects to form of either the local or global
variety, therefore it is useful to know how these objects interact 
with the dilaton.

In this paper we will consider the implications of superstring gravity 
for global cosmic strings. In particular we will address the question 
of whether a non-singular spacetime exists for the string. Recently 
Sen et.\ al.\ \cite{SEN} studied global strings in Brans-Dicke theory. 
We disagree with their claim that a static, non-singular solution 
exists for the string spacetime.
The structure of this paper is as follows. First, we 
briefly review the global string in Einstein gravity. 
We then study the global string in superstring gravity, for both 
massless and massive dilatons. Finally, we compare our 
results with the literature and present our conclusions - 
in particular, we present a novel solution which represents a 
self-gravitating string/anti-string pair on a closed 
two-dimensional spatial section.

\section{Non-static strings in Einstein gravity}

In this section, we review the global string in Einstein gravity. The Lagrangian
for an isolated $U(1)$ global string is 
\be
{\mathcal{L}} = \left ( \nabla_{a} \Phi \right ) ^{\dagger} \nabla^{a} \Phi - 
\frac{\lambda}{4} \left ( \Phi^{\dagger} \Phi - \eta^2 \right ) ^2 
\label{lagrangian}
\ee
By writing 
\be
\Phi = \eta X e^{i\chi} \label{phiform}
\ee
we reformulate the complex scalar field into two real, 
interacting scalar fields,
one massive ($X$), the other a massless Goldstone boson ($\chi$) 
\be
{\mathcal{L}}=\eta^2 (\nabla_a X)^2 + \eta^2 X^2 (\nabla_a \chi)^2 - 
\frac{\lambda \eta^4}{4} (X^2-1)^2
\ee
A vortex solution is characterised by the existence of closed loops in space
for which 
\be
\frac{1}{2\pi} \oint \frac{d\chi}{dl} dl = n \in \zed
\ee
We take the winding number $n=1$ and look for a solution describing an 
infinitely long, isolated, straight string.
The string spacetime is expected to exhibit cylindrical symmetry. If, in 
addition, we require the string to have fixed proper width, it can be shown
\cite{G} that coordinates may be chosen in which the metric is
\be
ds^2=e^{2A(r)} \left ( dt^2-e^{2b(t)} dz^2 \right ) -dr^2 - C^2(r) d\theta^2
\ee
where $\chi = \theta$, and $X=X(r)$.

Since we are dealing with an isolated global string, for convenience we set
$\sqrt{\lambda}  \eta =1$, choosing units in which the string width is $O(1)$.
Then the Einstein equations become 
\bea
A''+\frac{C''}{C} + A'^2 + \frac{A'C'}{C} & = & -\epsilon \hat{T}_t^t \\
(\ddot{b}+\dot{b}^2)e^{-2A}-A'^2-2\frac{A'C'}{C} & = & \epsilon \hat{T}_r^r 
\label{ein1} \\
(\ddot{b}+\dot{b}^2)e^{-2A}-2A''-3A'^2 & = & \epsilon \hat{T}_{\theta}^{\theta} 
\label{ein2}
\eea
where $\epsilon=8\pi G \eta^2$ measures the gravitational strength of the
string, and $\epsilon{\hat T}_{ab} = 8\pi GT_{ab}$. 
In addition, the field equation for $X$ is
\be
-X'' - \left (\frac{C'}{C}+2A' \right )X' + \frac{X}{C^2} + \half X (X^2-1)=0
\ee

Clearly (\ref{ein1}) and (\ref{ein2}) require 
\be
\ddot{b}+\dot{b}^2=b_0
\label{bzero}
\ee
where $b_0$ is a constant. Reference \cite{G} examined 
the behaviour of the string metric 
for different values of the free parameter $b_0$. This analysis is involved 
and will largely be mirrored in the next section; therefore 
we will merely sketch the results. The region of parameter space $b_0 \le 0$ is 
excluded as the metric can be proved to be singular in this case. For $b_0 > 0$ 
it can be shown that for a non-singular isolated string
solution to exist, $e^{2A} \rightarrow 
0$ at some finite $r_0$. This $r_0$ would represent an event horizon for the 
global string spacetime, analogous to that of the domain wall \cite{DW}. The 
asymptotic metric near this point would be
\be
ds^2 \simeq b_0 (r_0-r)^2 \left [ dt^2-\cosh^2 \sqrt{b_0} t dz^2 \right ] 
-dr^2-C_0^2 d\theta^2 \label{asymsol}
\ee
for some constant $C_0$. The curvature invariants for this metric are all 
finite, and a coordinate system can be found which extends beyond the event 
horizon $r=r_0$, verifying the coordinate nature of the singularity.

Of course, we have not shown such a non-singular solution necessarily exists. 
However, as demonstrated in \cite{G}, we can 
reduce the far-field equations describing 
the global string metric to a two-dimensional dynamical system. Whether or not 
(\ref{asymsol}) is admissible as an asymptotic solution for the global string 
reduces to the question of whether the dynamical system will asymptote the 
solution appropriate to (\ref{asymsol}) in phase space. By integrating out the 
full equations of motion to the edge of the vortex to find initial conditions 
for the dynamical system, it can indeed be shown that there always exists a 
$b_0$ for which a trajectory interpolates between the initial conditions and 
the asymptotic solution (\ref{asymsol}).

\section{Dilaton gravity and global strings}

We are interested in the behaviour of the global string 
metric when gravitational interactions take a form typical of low 
energy string theory. In its minimal form, string gravity replaces 
the gravitational constant, $G$, by a scalar field, the dilaton. 
To account for the unknown coupling of the dilaton to 
the global string, we choose the action 
\be
S=\int d^4 x \sqrt{-\hat{g}} \left [ e^{-2\phi} \left ( -\hat{R} 
- 4 (\hat{\nabla} \phi)^2 - \hat{V}(\phi) \right ) + e^{2a\phi} 
{\mathcal{L}} \right ] \label{stringaction}
\ee
where $\mathcal{L}$ is as in (\ref{lagrangian}). The potential for the
dilaton ${\hat V}(\phi)$ is for the moment assumed general.
This action is written in terms of the string metric which appears in the string
sigma model. To facilitate comparison with the previous section we instead
choose to write the action in the Einstein frame, which is related to the
string frame via the conformal transformation
\be 
g_{ab}=e^{-2\phi} \hat{g}_{ab}
\ee
in which the gravitational part of the action appears in the normal Einstein 
form
\be
S=\int d^4x \sqrt{-g} \left [ -R + 2(\nabla \phi)^2 - V(\phi) 
+ e^{2(a+2)\phi} {\mathcal{L}}(\Phi,e^{2\phi}g) \right ]
\ee
where $V(\phi)=e^{2\phi}\hat{V}(\phi)$. The energy-momentum tensor is now
\bea
T_{ab} &=& e^{2(a+2)\phi} \left [ \frac{\delta 
{\mathcal{L}}(\Phi,e^{2\phi}g)} {\delta g^{ab}} 
- \half g_{ab}{\mathcal{L}}(\Phi,e^{2\phi}g)\right ] \\
&=& {\eta^2\over 2} e^{2(a+1)\phi} \left [ 2\nabla_a X \nabla_b X + 2X^2
\nabla_a\theta \nabla_b\theta - g_{ab} \left 
( (\nabla X)^2 + X^2 (\nabla\theta)^2 -
{e^{2\phi}\over4} (X^2-1)^2 \right ) \right ] \nonumber 
\eea
where $\Phi$ takes the form (\ref{phiform}) and $\chi = \theta$ as before.
Einstein's equation becomes
\be
G_{ab}=  \epsilon{\hat T}_{ab} + S_{ab}
\ee
where ${\hat T}_{ab}$ is defined as before, and $\epsilon=\eta^2/2$. The second
term, $S_{ab}$, represents the energy-momentum of the dilaton 
\be
S_{ab}=2 \nabla_a \phi \nabla_b \phi +\frac{1}{2} g_{ab} V(\phi) 
- g_{ab} (\nabla \phi)^2
\ee
which has as its
equation of motion
\be
-\Box \phi=\frac{1}{4} \frac{\partial V}{\partial \phi} + \epsilon (a+1) 
\hat{T}_t^t + \quarter \epsilon e^{2(a+2)\phi} (X^2-1)^2
\ee
We will take a quadratic approximation $V(\phi)=2M^2 \phi^2$ (where the mass $M$
is measured in units of the Higgs mass) which will only be valid for small
$\phi$ when $M\neq 0$ -- an approximation which will require justification at
the appropriate stage. Note that we are ruling out a linear or exponential
potential which would correspond to a cosmological constant in the string
frame. In practise, we expect $10^{-12} < M < 1$, corresponding to a dilaton
mass in the range 1 TeV -- 10$^{15}$GeV. A dilaton with a lower mass than 1TeV 
would appear to be ruled out and a higher mass dilaton would give a {\it de
facto} Einstein gravitating global string.

As in Section 2, we choose the general, time-dependent, cylindrically
symmetric metric compatible with the symmetries of the source. That is
\be
ds^2=e^{2A(r)} \left ( dt^2 - e^{2b(t)} dz^2 \right ) -dr^2 - C^2(r)d\theta^2
\ee
where $b$ can be shown to satisfy (\ref{bzero}) as before. 
Taking $\sqrt{\lambda}\eta=1$, the rescaled, modified energy-momentum 
tensor is now
\bea
\hat{T}_t^t = \hat{T}_z^z & = & e^{2(a+2)\phi} \left ( e^{-2\phi} \left ( X'^2 +
\frac{X^2}{C^2} \right ) + \quarter (X^2-1)^2 \right )  \nonumber\\
\hat{T}_r^r & = & e^{2(a+2)\phi} \left ( e^{-2\phi} \left ( - X'^2 + 
\frac{X^2}{C^2} \right ) + \quarter (X^2-1)^2 \right ) \\
\hat{T}_{\theta}^{\theta} & = & e^{2(a+2)\phi} \left ( e^{-2\phi} \left ( X'^2 -
\frac{X^2}{C^2} \right ) + \quarter (X^2-1)^2 \right )\nonumber
\eea
and the equation of motion for $X$ is
\be
\left [ Ce^{2A} e^{2(a+1)\phi} X' \right ]' = \frac{X}{C} e^{2A}e^{2(a+1)\phi}+ 
\half Ce^{2A}e^{2(a+2)\phi} X(X^2-1) 
\ee
The gravi-dilaton equations are given explicitly by
\bea
\left [ C'e^{2A} \right ] ' & = & -\epsilon Ce^{2A}e^{2(a+1)\phi} \left ( 
\frac{2X^2}{C^2}+\quarter e^{2\phi}(X^2-1)^2 \right )-M^2 Ce^{2A}\phi^2 
\label{mfe1} \\
\left [ A'Ce^{2A}\right ]'& = & C \left ( b_0 -\quarter \epsilon e^{2A} 
e^{2(a+2)\phi} (X^2-1)^2 \right ) -M^2 Ce^{2A}\phi^2 \label{mfe2} \\
A'^2+\frac{2A'C'}{C} & = & b_0 e^{-2A} + \phi'^2-M^2\phi^2 -\epsilon 
e^{2(a+1)\phi} \left (  -{X'}^2 + \frac{X^2}{C^2} 
+\quarter  e^{2\phi}(X^2-1)^2 \right ) \label{mfe3} \\
\left [ Ce^{2A} \phi' \right ]' & = & M^2 Ce^{2A}\phi +\epsilon C e^{2A}
e^{2(a+1)\phi} \left (  (a+1) \left ( X'^2+\frac{X^2}{C^2} \right )
+   \frac{(a+2)e^{2\phi}}{4}  (X^2-1)^2 \right ) \label{mde}
\eea

Having set up the formalism and equations of motion,
we now turn to analyzing the possible solutions.

\subsection{Preliminary remarks}

We wish to examine under what conditions the spacetime of a global 
string may be non-singular. We consider in
turn the two cases $b_0 \le 0$ and $b_0 > 0$.

\subsubsection{$b_0 \le 0$}

We now prove that if $b_0 \le 0$ alll solutions are singular.
Note that (\ref{mfe1}) and (\ref{mfe2}) imply
\be
\left [ Ce^{2A} \right ]''=2b_0 C - \epsilon C e^{2A} e^{2(a+1)\phi} \left ( 
\frac{2X^2}{C^2}+\frac{3}{4}e^{2\phi} (X^2-1)^2 \right ) 
-3 M^2 Ce^{2A} \phi^2 \le 0 \label{star}
\ee
Thus $Ce^{2A} \le r$ for all $r$. Now if $(Ce^{2A})'=0$ at any point then 
(\ref{star}) implies $Ce^{2A}\rightarrow 0$ at some $r_0$ with $(Ce^{2A})'$ 
strictly negative. Then
\be
\frac{(Ce^{2A})'}{Ce^{2A}}=2A'+\frac{C'}{C} \rightarrow - \infty
\ee
as $r \rightarrow r_0$. Now (\ref{mfe2}) implies $A'$ is strictly negative away 
from $r=0$. So either $A'C'/C$ or $A'$ becomes infinite at $r_0$. But then
\be
R_{abcd}^2 \propto \left ( \frac{C''}{C} \right ) ^2 + 2 \left ( \frac{A'C'}{C} 
\right )^2 +2(A''+A'^2)^2 + (A'^2-b_0 e^{-2A})^2 \label{curv}
\ee
would become infinite at $r_0$ indicating a physical singularity. So for a 
non-singular spacetime, we take $(Ce^{2A})'>0$. 

Define
\bea
\alpha_1(r) & = & \epsilon \int_0^{r} \frac{e^{2A}e^{2(a+1)\phi} X^2}{C} dr
\nonumber \\
\alpha_2(r) & = & \epsilon \int_0^{r} \quarter Ce^{2A} e^{2(a+2)\phi} 
(X^2-1)^2 dr \nonumber \\ 
\alpha_3(r) & = & | b_0 | \int_0^{r} C dr\\
\alpha_4(r) & = & \epsilon \int_0^r Ce^{2A}e^{2(a+1)\phi}X^{\prime 2} dr
\nonumber \\
\alpha_5(r) & = & M^2 \int_0^r Ce^{2A}\phi^2 dr \nonumber
\eea
Then positivity of $(Ce^{2A})'$ implies
\be
1 > 2\alpha_1(r) + 3 \alpha_2(r) + 2 \alpha_3(r) + 3\alpha_5(r)
\label{relation}
\ee
for all $r$, hence the $\alpha_i$ (i = 1,2,3,5) are bounded. But
since $A'$ is negative and $(Ce^{2A})'$ positive, $C'$ must be greater
than zero for all $r$. Hence $\int C$ cannot converge and the spacetime 
must be  singular for $b_0 < 0$. 

For $b_0=0$, note that convergence of an $\alpha_i$ implies that its
integrand, $I_i$ must be o(${1\over r}$), so $Ce^{2A}I_i < rI_i \to 0$
as $r \to \infty$. Thus as $r\to\infty$ the constraint 
equation (\ref{mfe3}) becomes
\be
(\alpha_2+\alpha_5)^2 - 2(\alpha_2+\alpha_5)(1 - 2\alpha_1 - \alpha_2 -
\alpha_5) \simeq (Ce^{2A}\phi')^2 + Ce^{2A} I_4 > 0
\ee
Irrespective of the behaviour of $I_4$ and $\phi'$, we see that this
requires
\be
3(\alpha_2+\alpha_5) + 4\alpha_1 > 2
\ee 
Rearranging (\ref{relation}) shows that
\be
3(\alpha_2+\alpha_5) + 4\alpha_1 < 2 - 3(\alpha_2+\alpha_5)
\ee
Hence we have a contradiction, and are forced to conclude that whether the
dilaton is massless or massive, no non-singular 
solution exists for $b_0 \leq 0$.

Note that this argument does not use the specifics of the global string 
source, it simply uses the negativity of $A'$. This is a consequence of
the fact that ($T^r_r+T^\theta_\theta$) is positive, a general feature
of global defects. 

\subsubsection{$b_0 >0$}

For $b_0>0$, $(A'Ce^{2A})'$ no longer has a definite sign and
the arguments of the previous subsection will not work. We start by noting
that (\ref{mfe1}) implies $(C'e^{2A})'\le 0$ and hence $C'e^{2A}\le 1$. 
Either $C'e^{2A}$ remains positive, or it does not. However, if
$C'e^{2A}>0$ for all $r$, then $\alpha_3(r)$ will diverge, implying that
$e^{2A}\to\infty$ as $r\to\infty$. An examination of (\ref{mfe1},\ref{mfe2})
shows that
\be
C \to C_0 \;\;\; , \;\;\;\; e^A \sim \sqrt{b_0}r \;\;\; {\rm as} \;\;\;
r\to\infty
\nonumber
\ee
If $M^2\neq0$, then finiteness of $\alpha_5$ requires $\phi \to 0$ at 
large $r$, and finiteness of  $\alpha_1$ and $\alpha_2$ then requires
$X\to 0$ and $X\to 1$ respectively - clearly an impossibility. If $M^2 = 0$,
then finiteness of $\alpha_1$ and $\alpha_2$ requires
\bea
e^{2(a+1)\phi}X^2 & = & o(1/r^3)\nonumber \\
e^{2(a+2)\phi} (X^2-1)^2 & = & o(1/r^3)\nonumber
\eea
and so either $e^{2(a+1)\phi}=o(1/r^3)$ or $e^{2(a+2)\phi}=o(1/r^3)$ 
or both. Thus 
$|\phi|\rightarrow \infty$ as $r\rightarrow \infty$. But 
(\ref{mde}) shows that either $\phi$ is bounded or $(a+1)\phi\to\infty$,
hence $X^2, e^{2(a+2)\phi} = o(1/r^3)$ as $r\to\infty$. But these are
contradictory statements, since if $X$ is so bounded, the integral
$\alpha_4$ involving $X'$ can never diverge.
Thus at some $r_0$, $C'e^{2A}=0$. 

Suppose that $C'=0$ but $e^{2A}\ne 0$. Then (\ref{mfe1}) implies $C''<0$, 
so $C'$ becomes negative and must remain negative. Now if 
$C \to 0$ at $r_1$, say, 
then non-singularity of the spacetime from (\ref{curv}) 
requires that $C'', A' \to 0$. If we require the global string to be 
isolated, $X \approx 1$, and hence from (\ref{mfe1}), $e^{2(a+1)\phi} \to 0$. 
But then integrating (\ref{mde}) shows $Ce^{2A}\phi'=O(r_1-r)^2$ 
which means that $\phi(r_1)$ is bounded - a contradiction. If we 
drop the requirement that $X \approx 1$ then we have the interesting 
possibility that $X \to 0$ and there is an anti-string at $C=0$. 
We will explore this further in the final section.

For a non-singular, isolated string solution, we require $e^{2A}\to 0$ 
as $r \to r_0$. Near this point the asymptotic solution for the metric is 
\bea
e^A & \sim & \sqrt{b_0} (r_0-r) \nonumber \\
C &=& C_0 + O(r_0-r)^2 \label{asymetric}
\eea
which is the event horizon discussed in \cite{G}.
The value of $b_0$ may be implicitly determined via integration of
(\ref{mfe2}) which yields
\be
\alpha_3 = \alpha_2 + \alpha_5
\label{bfix}
\ee
for which a linearized argument would give $b_0 = O(\epsilon e^{1/\epsilon})$,
but a more general argument \cite{G} can certainly bound $b_0$ by
$\epsilon^2$.

If the dilaton is massless we can integrate the dilaton 
equation (\ref{mde}) out to $r_0$ and obtain
\be
Ce^{2A}\phi' |_{r_0} = (a+1) \left ( \alpha_1(r_0) + \alpha_4(r_0) \right )
+(a+2) \alpha_2(r_0) 
\label{dilint}
\ee
Since nonsingularity of $R_{ab}^2$ requires that $\phi'$ is finite at $r_0$, 
the LHS is zero and hence $a \in (-2,-1)$. 

Of course, so far, we have merely restricted the parameter ranges in which 
it may be possible to find a non-singular solution. That is, $b_0>0$, 
and $-1>a>-2$ if $M^2=0$. It remains to be shown that such a 
solution exists. We will do this by reducing the far-field 
equations to a two-dimensional dynamical system and demonstrating 
the existence of a trajectory interpolating between the initial 
conditions at the edge of the string and the asymptotic solution 
given by (\ref{asymetric}).

\subsection{Solutions for a massless dilaton}
We take $X=1$ outside the core and look for the asymptotic solution. 
Our far-field equations are
\bea
\left [ C'e^{2A} \right ]' & = & -\frac{2\epsilon 
}{C}e^{2A}e^{2(a+1)\phi} \nonumber\\
\left [ A'Ce^{2A} \right ]' & = & b_0 C \nonumber \\
A'^2+\frac{2A'C'}{C} & = & b_0 e^{-2A} + \phi'^2 
-\frac{\epsilon e^{2(a+1)\phi}}{C^2} \label{farfe} \\
\left [ Ce^{2A}\phi' \right ] ' & = & \frac{\epsilon 
}{C}(a+1)e^{2A}e^{2(a+1)\phi} \nonumber
\eea
We will take $\epsilon\ll 1$ and $b_0<O(\epsilon^2)$. 
Let $\rho= \int_0^r e^{-A}dr$ and denote $\frac{d}{d\rho}$ by a dot. 
Then letting $f=\dot{A}+ \frac{\dot{C}}{C}$, $g= \frac{\dot{C}}{C}$ 
and $h=\dot{\phi}$, some manipulation of the far-field equations gives 
\bea
\dot{f} & = & f^2-2g^2-2h^2-b_0 \nonumber \\
\dot{g} & = & -2(b_0+g^2+h^2-f^2)-fg\\
\dot{h} & = & (a+1)(b_0+g^2+h^2-f^2) -fh \nonumber 
\eea
Note that
\be
(a+1) {\dot g} + 2 {\dot h} + f \left [ (a+1)g + 2h \right ] = 0
\ee
Hence since both $h$ and $g$ are zero at $r_0$, 
\be
h=-\frac{a+1}{2}g
\label{hgpro}
\ee
at all points on the trajectory corresponding to a physical global string.
Integrating this relation gives
\be
e^{2\phi} \propto C^{-(a+1)}
\label{massless}
\ee
for the non-singular solution.

Therefore on the plane given by (\ref{hgpro}) our far-field equations reduce 
to the two-dimensional dynamical system
\bea
\dot{f} & = & f^2-2\gamma g^2-b_0\nonumber \\
\dot{g} & = & 2f^2-2\gamma g^2 -2b_0-fg \label{twods}
\eea
where
\be
\gamma= \left ( 1+\frac{(a+1)^2}{4} \right )
\ee
For the parameter range we are interested in, $1<\gamma<1.25$.
However, the relation (\ref{hgpro}) between $h$ and $g$ will determine
the specific value of $\gamma$ from the matching
conditions at the string core.

\begin{figure}
\begin{center}
\epsfig{figure=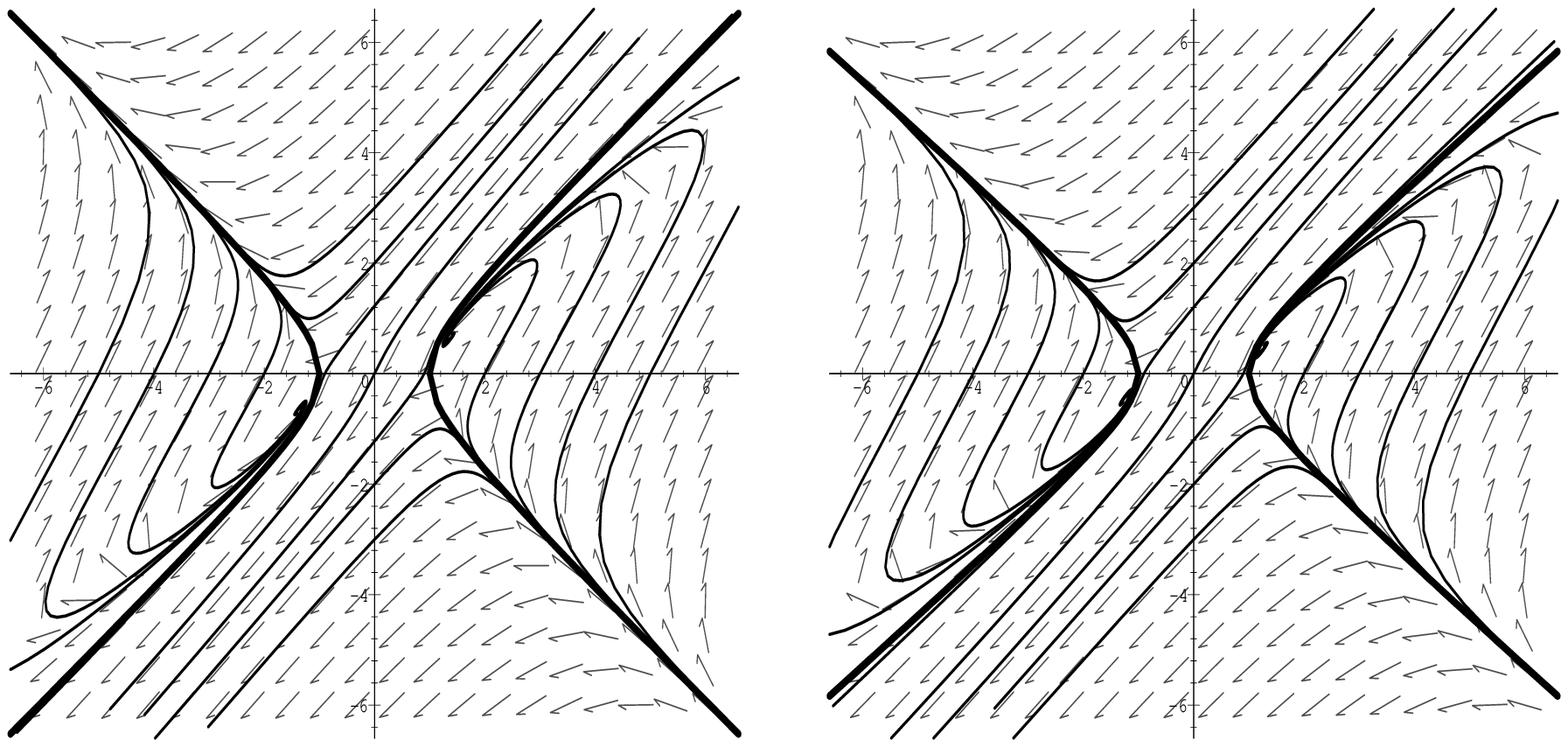,height=8cm}
\caption{The phase diagram ($b_0>0$) with $\gamma=1$ and $\gamma=1.25$.
\label{fig:plot1}}
\end{center}
\end{figure}

Consider $b_0>0$. We set $t=\sqrt{b_0}\rho$ and $(f,g)=\sqrt{b_0}(x,y)$. Then
\bea
\frac{dx}{dt} & = & x^2-2\gamma y^2 -1 \nonumber\\
\frac{dy}{dt} & = & 2x^2-2\gamma y^2 -2 - xy
\eea
The system has fixed points at
\be
(\pm 1,0), \;\;\;\;
\left ( \pm \sqrt{\frac{2\gamma}{2\gamma-1}} , 
\pm \frac{1}{\sqrt{2\gamma (2\gamma-1)}} \right )
\ee
These are, respectively, saddle points and foci. A diagram 
of the phase plane for $\gamma=1$ and $\gamma=1.25$ is 
given in Figure \ref{fig:plot1}.

Consider the fixed point $(-1,0)$. This corresponds to 
$f=-\sqrt{b_0}, g=0$. That is 
\be
(e^A)'=-\sqrt{b_0}, C'=0
\ee
the asymptotic form of the metric (\ref{asymetric}). The question 
of whether a non-singular solution exists for the global string 
reduces to asking whether a suitable trajectory exists terminating on 
the critical point $(-1,0)$. Since this is a saddle point, there 
is a unique trajectory approaching $(-1,0)$, the stable manifold. 
We must examine whether this trajectory matches onto the core of the string.

Let $\rho_c$ be a suitable value of $\rho$ representing the 
edge of the string and let $r_c$ be the corresponding value of $r$. 
Then from (\ref{mfe2})
\be
A'Ce^{2A}|_{r_c}=b_0 \int_0^{r_c}C dr - \epsilon \int_0^{r_c} 
\quarter Ce^{2A}e^{2(a+2)\phi} (X^2-1)^2 dr
\ee
Assuming $b_0<O(\epsilon^2)$ then 
\be
A'e^A \approx -\frac{\epsilon}{Ce^A} \int_0^{r_c} {\quarter} C 
e^{2A} e^{2(a+2)\phi} (X^2-1)^2 dr \approx -\frac{\epsilon K_2}{Ce^A}
\ee
where 
\be
K_2=\int_0^{r_c} {\quarter} Ce^{2A} e^{2(a+2)\phi} (X^2-1)^2 dr = 
{\alpha_2(r_c) \over \epsilon}
\ee
But $e^A\approx 1$, $C\approx r_c$ and $\rho_c=\int_0^{r_c}e^{-A}dr 
\approx r_c$. Hence
\be
\dot{A}(\rho_c) = (e^A)'|_{r_c} \approx - \frac{\epsilon K_2}{\rho_c}
\ee
Similarly, from (\ref{mfe1}), we obtain
\be
\frac{\dot{C}}{C}(\rho_c) \approx \frac{1}{\rho_c}(1-\epsilon (K_2+2K_1))
\ee
where
\be
K_1=\int_0^{r_c} \frac{e^{2A}e^{2(a+1)\phi}X^2}{C} dr = 
{\alpha_1(r_c) \over\epsilon}
\ee
Hence
\be
f_0 \approx \frac{1}{\rho_c}(1-2\epsilon (K_1+K_2))
\ee
and
\be
\frac{g_0}{f_0}=\frac{y_0}{x_0}=\frac{1-\epsilon (K_2+2K_1)}
{1-2\epsilon (K_1+K_2)} \approx 1+ \epsilon K_2
\ee

This gives an initial relation for $f$ and $g$. However, recall that 
in order to get a two-dimensional dynamical system we eliminated
$h$ (the dilaton) from consideration by projecting onto the plane given by
(\ref{hgpro}).  Clearly this relation holds all along 
the stable manifold and hence in 
particular must hold at the point at which the non-singular trajectory matches
to the string core. But
\bea
h(\rho_c) &=& {1\over Ce^A} \left [ (a+1)(\alpha_4+\alpha_1) +
(a+2)\alpha_2 \right ]_{\rho_c} \nonumber \\
&=& {1\over \rho_c} \left [ (a+1) (\alpha_4(r_c) + \epsilon K_1) +
(a+2)\epsilon K_2 \right ]
\eea
thus
\bea
2h(\rho_c) + (a+1)g(\rho_c) &=& (a+1)(1+O(\epsilon)) + (a+3) \epsilon K_2 = 0
\nonumber \\
\Rightarrow \hskip 1cm
(a+1) &=& -2\epsilon K_2 + O(\epsilon^2)
\eea
or $\gamma = 1 + O(\epsilon^2)$. Therefore the matching of the
asymptotic trajectory onto the core requires that $a$ is very close
to $-1$.

Therefore the trajectory approaching $(-1,0)$ in the $(x,y)$ plane will 
correspond to a global string if it intersects the line 
$y=(1+K_2 \epsilon)x$ for some $x>0$ for the specific value of
$\gamma$ given by the initial conditions. 

Now
\be
\frac{dy}{dx}=1+\frac{x^2-xy-1}{x^2-2\gamma y^2-1}
\ee
We can see that on the line $y=x+1$, $dy/dx>1$ and on the line 
$y=2(x+1)$, $dy/dx<2$. Since by observation $y \in [1,2]$ at $x=0$ 
for the non-singular trajectory, we can bound $y$ by $[x+1,2(x+1)]$ 
for general $x>0$. Now as $t\rightarrow -\infty$, $x,y\rightarrow 
\infty$ and we have
\be
\frac{dy}{dx} \approx 1+ \frac{x^2-xy}{x^2-2\gamma y^2}
\ee
Suppose $\gamma=1$. Then to leading order, the solution for $x$ 
and $y$ can be written 
\be
y \approx x \left [ 1+ \frac{1}{4\ln x} \right ]
\ee
For non-zero $\epsilon$, there is an $x_{\epsilon}$ such 
that $1+1/4\ln x < 1+ K_2 \epsilon$ for all $x>x_{\epsilon}$ 
and hence the trajectory will intersect $y=(1+K_2 \epsilon)x$ 
at some value of $x$. But the presence of $\gamma > 1$ merely 
makes $dy/dx$ on the trajectory even smaller. Thus for any 
$1<\gamma<1.25$ the trajectory definitely intersects 
$y=(1+K_2 \epsilon)x$, therefore in particular, for $\gamma = 1 + \epsilon^2
K_2^2$, the trajectory matches on to a core solution. 
This demonstrates the existence 
of a non-singular solution for the global string metric 
for a certain value of $b_0$.

\subsection{Massive dilatonic gravity}

We now carry out a similar analysis for massive dilatonic gravity. 
In this case, for mid-range values of $r$, we can use linearised theory 
to integrate the dilaton equation (\ref{mde})
to get
\bea
\phi(r) & = & -\epsilon K_0(Mr) \int_0^r I_0(Mr') r'
\left [ (a+1) \left ( X^{\prime2} + {X^2\over r^{\prime2}} \right )
+ {(a+2)\over 4} (X^2-1)^2 \right ] dr' \nonumber \\
& & - \epsilon I_0(Mr) \int_r^\infty K_0(Mr') r'
\left [ (a+1) \left ( X^{\prime2} + {X^2\over r^{\prime2}} \right )
+ {(a+2)\over 4} (X^2-1)^2 \right ] dr' \\
& & \simeq - {\epsilon (a+1) \over M^2 r^2} \nonumber
\eea
We also see that $\phi(0) \simeq -\epsilon (a+1) (\ln M)^2$. Therefore, as
opposed to the local string\cite{GS} or the global monopole\cite{DG},
$|\phi(0)|$ is more strongly varying with $M$, proportional to the square,
rather than the magnitude of the logarithm. However, for a GUT scale 
defect, $\epsilon = 10^{-6}$, and $|\ln M |< 28$, therefore we see that 
$\phi$ is still just small enough to justify the quadratic approximation.

Outside the core the far field equations are
\bea
\left [ C'e^{2A} \right ] ' & = & -\frac{2\epsilon e^{2A}}{C}\label{far1} \\
\left [ A'Ce^{2A} \right ] ' & = & b_0 C\label{far2}\\
A'^2+\frac{2A'C'}{C} & = & b_0 e^{-2A} - \frac{\epsilon }{C^2}\label{far3}\\
\left [ Ce^{2A} \phi' \right ] '& = & M^2 Ce^{2A} \phi + 
\frac{\epsilon (a+1) e^{2A} }{C} \label{dill}
\eea
hence we see that the correct leading order solution for $\phi$ outside the
core allowing for the geometry is
\be
\phi \simeq - \frac{\epsilon (a+1) }{M^2 C^2}
\label{massive}
\ee
for $a\neq-1$ and 
\be
\phi \simeq - {\epsilon\over M^2 C^4}
\label{massivea}
\ee
for $a = -1$.
(\ref{far1}-\ref{far3}) are identical to their Einstein gravity
counterparts. We may therefore use the arguments of \cite{G} (or the
previous subsection for $\gamma=1$) to deduce the existence of a solution.

\section{Discussion}
 
We have studied the behaviour of the metric and dilaton field 
of a global cosmic string in superstring gravity for an arbitrary 
coupling of the string Lagrangian to the dilaton : $e^{2a\phi} 
{\cal L}$. For both massless and massive dilatons, we have 
demonstrated the existence of a non-singular spacetime for 
the string if we include an exponential expansion along the length
of the string $e^{b_0t}$, with $b_0 > 0$. In addition we have 
the further restriction that $a=-1 - 2 \epsilon K_2$ 
for the massless dilaton. In both cases, the spacetime is characterised 
by an event horizon at finite distance from the string core. 
Near this point, the asymptotic solution for the metric is 
\be
ds^2 \approx b_0 (r_0-r)^2 \left [ dt^2 -\cosh^2 \sqrt{b_0}t 
dz^2 \right ] - dr^2 - C_0^2 d\theta^2
\ee
However, since this non-singular solution is very similar to the Einstein 
global string, we expect that it too will be unstable.
The metric at intermediate points will be given by the Cohen-Kaplan
\cite{GSTR} solution, and the dilaton by either (\ref{massless}) if it is
massless, (\ref{massive}) if it is massive and $a\neq-1$, or
(\ref{massivea}) if the dilaton is massive and $a =-1$.
Thus we expect that the cosmological effects of global strings 
deriving from their purely gravitational or metric properties \cite{HH}
will be little altered from the Einstein case. The main difference will be
dilaton production by global strings. In this case, as with the global
monopole \cite{DG}, the Damour-Vilenkin \cite{DV} bounds on the 
dilaton mass hold for $a\neq-1$, but if a global string couples to
a massive dilaton with $a = -1$, then the Damour-Vilenkin bound is
weakened slightly. For example, for a TeV mass dilaton, the Damour-Vilenkin
limit on the symmetry breaking scale $\eta$ is 10$^{13}$ GeV, which is
weakened to 10$^{14}$ GeV in the case $a = -1$.

In the course of our analysis, we have shown 
that the spacetime of a static ($b_0=0$) global string 
in dilaton gravity is necessarily singular. This is in 
disagreement with the work of Sen {\it et.\ al.\ } \cite{SEN} who 
studied static global strings in Brans-Dicke theory, 
apparently finding non-singular solutions to the far-field equations. By 
utilising the dynamical system approach, we can see that while 
quite valid as solutions to the far-field equations, these are not 
solutions to the full field equations.
 
\begin{figure}
\begin{center}
\epsfig{figure=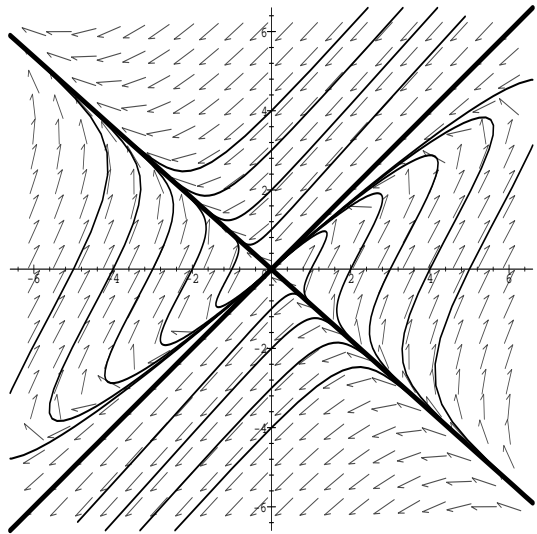,height=8cm}
\caption{The phase diagram for $b_0=0$ with $a=0$.
\label{fig:plot2}}
\end{center}
\end{figure}

Massless dilatonic gravity corresponds to Brans-Dicke theory 
for the particular parameter values $\omega=-1$ and $a=0$. 
With $b_0=a=0$ the far-field dynamical system (\ref{twods}) is
\bea
\dot{f} & = & f^2 - \frac{5}{2}g^2 \\
\dot{g} & = & 2 f^2 - \frac{5}{2} g^2 - fg
\eea
with a single fixed point at $f=g=0$ (see Figure \ref{fig:plot2}). The 
separatrices are $f-g$ 
and $f+ \frac{\sqrt{5}}{2}g$. The non-singular solution found by 
Sen {\it et.\ al.\ } corresponds to
\be
f=g =  \frac{2}{3(r-r_0)}
\ee
where $r_0 \le 0$. This is
the part of the separatrix $f=g$ lying in the upper-right 
quadrant of the phase plane with the flow towards the 
origin. However, $f-g = {\dot A}$, which is zero at the core of the 
global string, and integrating (\ref{mfe2}) shows
\be
(f-g) Ce^A = - {\epsilon\over4} \int_0^\rho
Ce^{3A} e^{4\phi} (X^2-1)^2 d\rho <0
\ee
i.e.\ $g$ is strictly greater than $f$ outside the core. Thus the
trajectory $f=g$ cannot correspond to the exterior of a physical
global string.

Finally, we would now like to comment on the 
possibility discussed earlier, namely that $C\to 0$ 
at some finite $r_1$, giving a solution on a compact $(r,\theta)$ 
section. Since the field equations are symmetric under the 
transformation $r \to r_1-r$, their solution must also be 
symmetric. Thus at $r_0= r_1/2$, we have $C'=A'=\phi'=0$. Defining $f,g$ 
and $h$ as before, this means that
$f=g=h=0$ at $r_0$. Whether the dilaton is massive or massless,
we can reduce the far-field 
equations to the same two-dimensional dynamical system (\ref{twods}). 
The symmetric
solution must correspond to the trajectory going through the origin,
since $f=g=0$ at $r_0$. For $f,g>0$ this trajectory is trapped between 
the non-singular isolated string trajectory and the line $y=(1+K_2 \epsilon)x$. 
Hence, we can again argue that for some $b_0>0$ the solution 
represented by this trajectory matches onto the core of the global 
string near $r=0$. For $f,g<0$, we can similarly argue that as $r\to r_1$ 
and $f,g\to -\infty$, the trajectory matches onto the core of 
an anti-string located at $r_1$. 

\begin{figure}
\begin{center}
\epsfig{figure=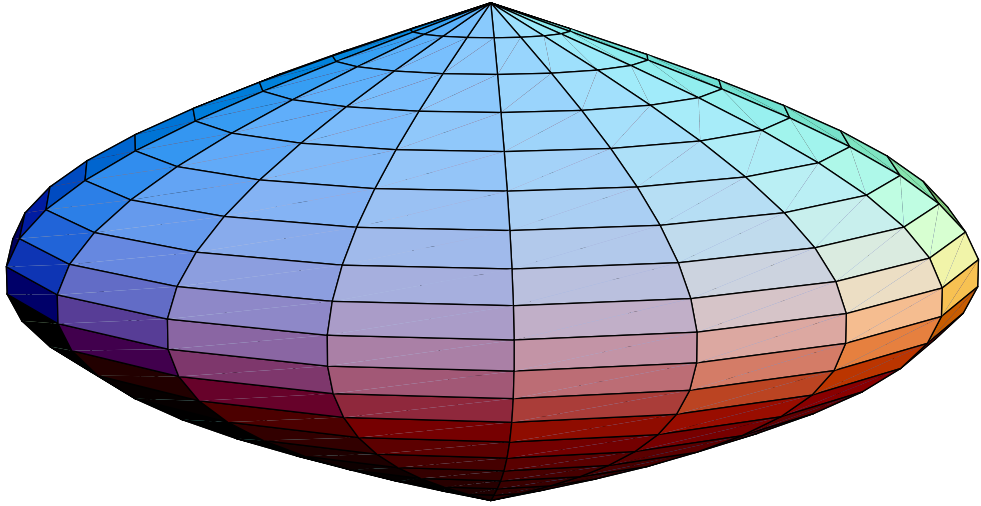,height=8cm}
\caption{A schematic representation of the $(r,\theta)$ spatial section
for a string/anti-string pair.
\label{fig:plot3}}
\end{center}
\end{figure}

The $(r,\theta)$ spatial sections of this spacetime 
are compact, and are qualitatively depicted in Figure \ref{fig:plot3},
with the string and anti-string at opposite poles.
We can estimate the scale at which this compactification occurs by 
integrating (\ref{mfe1}), which gives $r_0$ of the order of $e^{1/\epsilon}$.
For $\epsilon \sim 10^{-6}$, this is way beyond the current cosmological
horizon, however, it is tempting to extrapolate this solution to larger
values of $\epsilon$. If $\epsilon$=O(1), then the $(r,\theta)$ sections
would be compact at a scale of order $\epsilon$, and the spacetime would
be effectively two-dimensional with an exponential expansion in the 
spatial dimension. Of course this is way beyond the validity of our analysis,
however, it is interesting to consider in the light of topological
inflation scenarios \cite{TOP}. In these, a Planck scale topological defect
is considered as a source for inflation. If this string/anti-string
solution persists at high energy, then the global string would not be
a suitable candidate for topological inflation.

\vskip 1cm
\noindent{\it Note added}

After completing this work, we note that Boisseau and Linet
\cite{BL} have recently computed the exterior metric of
a global string in Brans-Dicke gravity. 

\vskip 2cm

\section*{Acknowledgements}
 
\centerline{ O.D.\ is supported by a PPARC studentship,
and R.G.\ by the Royal Society.}

\end{document}